# Teacher-Driven Professional Development and the Pursuit of a Sophisticated Understanding of Inquiry

Mike Ross, Ben Van Dusen, Samson Sherman, Valerie Otero

*University of Colorado, 249 UCB, Boulder, CO 80309*

**Abstract.** The need for highly qualified physics teachers in the U.S. is well established, and reform efforts are underway to develop novel and innovative teacher professional development experiences to improve the quality of K-12 physics education. Streamline to Mastery is an NSF-funded, learner-centered professional development program that seeks to capitalize on teachers' knowledge and experience to move physics teachers toward mastery in their fields. Teacher participants in this teacher-driven program choose their own goals and areas of growth. One of these areas has been the development and implementation of inquiry-oriented curriculum, as well as the adaptation of traditional lessons toward a greater inquiry orientation. Results indicate that teachers' conceptions of inquiry teaching and learning have become more expert-like as they have engaged in teacher participant-driven experiences in the pursuit of greater understanding and more effective classroom practice.

**Keywords:** Inquiry, Teacher professional development, Teacher-Driven

**PACS:** 01.40.jh

## INTRODUCTION

Streamline to Mastery is an NSF-funded, teacher-driven physical science teacher professional development program. The goals of Streamline to Mastery are to support teachers in their endeavors to improve their own professional practices and to develop a community of science education leaders within the greater population of practicing science teachers. These are the only formalized learning goals of this professional development program, and further and more specific goals must, by design, emerge from the teachers' own perceived needs and areas of interest.

The first cohort of four Streamline to Mastery teachers chose to focus on the topic of inquiry-oriented science instruction, and this has been a dominant thread in the first year of the professional development program. As other researchers have noted, the term *inquiry* in science education has somewhat of a troubled past in that teachers and researchers alike often characterize inquiry teaching and learning in disparate ways [1]. Researchers have investigated pre-service teachers' conceptions of inquiry and found them to be inconsistent with those of practicing scientists [2], and others [3] have found experienced science teachers' conceptions of classroom scientific inquiry to be incomplete as compared to NRC published documents such as *Inquiry and the NSES* [4]. However, the circumstances that give rise to these ambiguities and lack of shared understanding are less clear. This study attempts to shed light on these issues by examining the longitudinal trajectories of a small sample of practicing science teachers as they define classroom scientific inquiry for themselves through reflective practice and community discourse under the guidance of physics education researchers.

## RESEARCH CONTEXT

Four middle and high school physical science teachers from urban schools participated in this study. These schools both have large populations of English language learners, and the majority of students in these schools qualify for the free or reduced lunch program. These teachers were recruited as the first of two cohorts of secondary physical science teachers from high needs schools to participate in Streamline to Mastery for five years. As shown in Table 1, all but one of the four teachers has been teaching three years or less and three of the four are teaching outside of their scientific discipline. Two of the four teachers are also former university Learning Assistants [5].

**TABLE 1.** Participant Demographics

| Degree | Years Exp. | Subject Taught |
|---|---|---|
| B.A.Bio/Ph.D. Biochem | 1 | Physics |
| B.A. Chem/M.A. Urban Ed | 3 | Physics |
| B.A. Phys/M.A. Urban Ed | 3 | Physical Sci. |
| B.A. Bio/M.A. Urban Ed | 7 | Physical Sci. |

Requirements to be in the program included teaching in a high needs district, completion of a master's degree, and a willingness to share aspects of teaching practice and collaborate. Additionally, teachers are required to conduct research into their own practices, present at least once per year at a national education conference, and take one graduate level college course of their choice per year for credit. The research team, all of whom participated directly in the program, consisted of the NSF project PI, two doctoral students in physics education research who were formerly high school physics teachers, and one future physics teacher who is currently serving as a Noyce Fellow.

Teachers and researchers met semi-weekly to share lessons, plan classroom research, and discuss topics of interest to the teachers. Activities included lesson-sharing in which teachers and researchers each shared a lesson that they deemed to be effective and inquiry-oriented with the other teachers and researchers. The teacher participants also attended numerous national conferences, including one in which they presented a poster on the Streamline to Mastery program and another in which they collectively led a workshop on inquiry-oriented science instruction.

## METHODS

The data collected for this study consists of lesson sharing reflections, responses to two administrations of prompts taken from a survey of teachers' conceptions of inquiry, and video of professional development meetings. The community itself was the unit of analysis. These data were used to triangulate our findings about teachers' understanding of inquiry and were analyzed using the five essential features of inquiry specified in the NRC document *Inquiry and the National Science Education Standards: A guide for teaching and learning* [4]. These five essential features and their hereafter abbreviated names are (1) Engaging in scientifically oriented questions (Questions), (2) Giving priority to evidence (Evidence), (3) Formulating explanations based on evidence (Explain), (4) Evaluating explanations in connection with scientific knowledge (Connections), and (5) Communicating explanations (Communicate). Though the research team does not assert that the NSES description of inquiry should or does represent a "gold standard" for inquiry-oriented instruction, we chose to employ it as a framework for assessing teachers' understanding of inquiry and consider conceptions consistent with this framework to be "expert-like" for the purposes of this study. It should be noted that the researchers never made this framework or other related literature available to the teacher participants.

### Inquiry Survey Item Responses

The inquiry survey used in this study was designed by Kang and Wallace [3] to assess secondary science teachers' conceptions of inquiry relative to the five essential features of inquiry. As an example, the scenario "Giving students a white powder" was designed specifically to elicit a response related to the inquiry feature *Giving priority to evidence* (Evidence). Rather than using the survey items as the designers intended, we used the items to cue extended open-ended responses about the topic of inquiry more generally. These response data were coded for the five essential features as well as other notable response patterns not captured in the NSES inquiry framework.

### Lesson-sharing Reflections

In addition to survey item responses, the participants also generated reflections following each of the five lesson sharing events. As stated above, the teachers and researchers each chose a lesson that they deemed to be effective and inquiry-oriented to teach to the group. After the lesson sharing, teachers and researchers debriefed aspects of the lessons together and teachers completed one online lesson-sharing reflection for each lesson. The five lesson-sharing reflections were responses to the prompts: (1) *In what ways was this an inquiry lesson?* and (2) *How might you modify this lesson for your classroom?* These were recorded using an online message board and participants could see the posts made by the others.

The five lesson-sharing reflections occurred in the first five months of the study, and, because inquiry was a recurring topic of teacher discussion and concern throughout the first year, the inquiry survey items were administered both in the $10^{th}$ and $17^{th}$ month of the project to assess teacher conceptions. These administrations allowed us to gain longitudinal data on teacher conceptions.

## FINDINGS

At the start of the study, teachers' lesson sharing reflections and meeting discourse were ambiguous with regard to the subject of inquiry. The term *inquiry* was frequently used by teachers in a manner that made it difficult for the researchers to distinguish its intended meaning from other terms used frequently such as *constructivism, hands-on, real-world,* and even *best practices* based on the context of the conversations and reflections. For example, when responding to the lesson-sharing reflection prompt asking what made the shared lessons inquiry-oriented, teachers offered these responses:

**Teacher 1:** *"…it really focused on kids trying to figure it out for themselves. A true hands-on activity."*
**Teacher 2:** *" As a result of the process and labs, students construct an understanding of how carriers…"*
**Teacher 3:** *"This [name of curriculum] really does guide students' thinking through an abstract concept starting with a real-life application…"*

Teacher 2 also used the terms *inquiry* and *constructivism* interchangeably in meeting conversations: *"How can I use already created materials while upholding a constructivism or inquiry approach in the classroom?"* Teacher 4 used terminology ambiguously as well, but offered expert-like responses to the lesson-sharing prompt following the first lesson share: *"Students were really doing science by testing their model and revising it based on experimental evidence"*.

As shown in Figure 1, the frequency with which the teacher participants characterized the shared lessons' inquiry-orientation using the five essential features of inquiry increased over the course of the five lessons. In lessons 1 and 3, only *giving priority to*

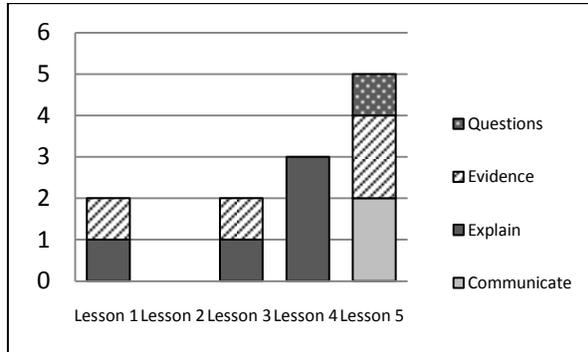

**FIGURE 1.** Frequency of References to Five Essential Features of Inquiry in Lesson Reflections

*evidence* (Evidence) and *formulating explanations based on evidence* (Explain) were noted. In lesson 2, no references were made that aligned with the five essential features of inquiry. By the fifth lesson, however, four of the five essential features were referenced. Only *evaluating explanations in connection with scientific knowledge* (Connect) was not referenced. These data show increasing numbers of references consistent with the NSES inquiry framework over the course of the five lessons and suggest evolution of the teachers' conceptions of inquiry toward a more expert-like understanding. It should also be noted that the coded references were relatively evenly distributed across the four participants over each of the data sources.

References not captured by the five NSES essential features of inquiry were observed and coded, as they were deemed relevant to characterizing these teachers' conceptions of inquiry and appeared frequently in other data sources. As shown in Figure 2, references to the Real-world oriented tasks (Real-world) were common as were references to the social nature of student activities (Social), ownership of ideas, tasks, procedures, etc. (Own), scientific models (Model), and constructivist epistemology (Constructivism).

The inquiry survey data were coded using the same system developed for the lesson-sharing reflections and, though the survey was identical in the two administrations, the results bore some notable differences. As shown in Figure 3, the frequency with which teachers referenced *Communicate* and *Explain* increased approximately five-fold for each category, and the frequency with which they referenced *Evidence* stayed roughly the same. As with the lesson-sharing reflection data, no references to *Connect* were made. The increase in the number of references that were consistent with the NSES inquiry framework is further evidence of these teachers' evolution toward a more expert-like understanding of inquiry.

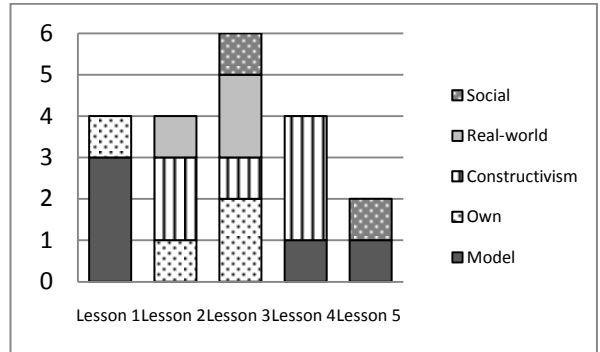

**FIGURE 2.** Frequency of References Not Captured in the NSES Inquiry Framework

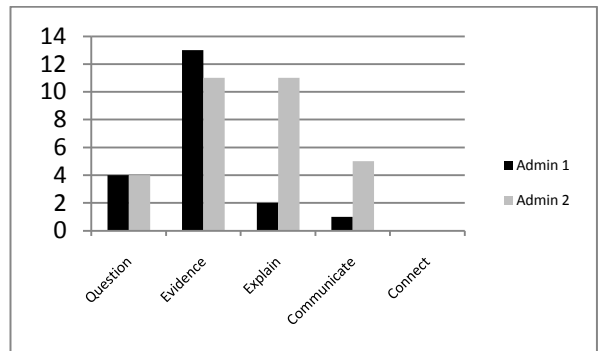

**FIGURE 3.** Frequency of References to Five Essential Features on Inquiry Survey Items

As is shown in Figure 4, references to *Real-world* and *Model* decreased markedly from the first to second administrations of the inquiry survey, and references to *Social* decreased by half. References to *Constructivism* and *Own* increased moderately. Though some variation is noted across the two administrations, it is clear from the data that these teachers feel that ownership of ideas, social construction of knowledge, and real-world relevance are also key features of inquiry.

Finally, it is important to note that in the 12[th] month of the study, Teacher 2 motivated a discussion with the aim of defining inquiry. Teacher 2 had attended a conference and engaged with colleagues in a conversation about inquiry learning. Upon returning

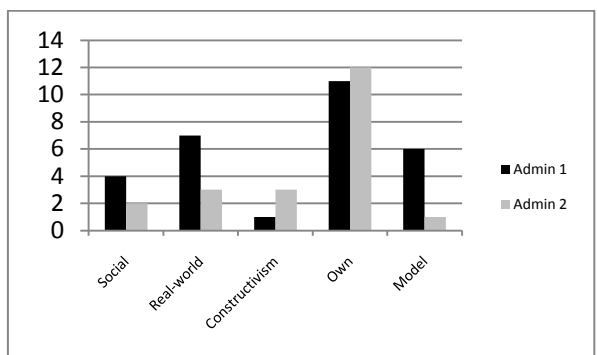

**FIGURE 4.** Frequency of References Not Captured in the NSES Inquiry Framework

Teacher 2 shares:*"I felt like I should have a more cohesive ability to discuss it, or more cohesive description of inquiry, with everything that we've done, and I felt like I was somewhat articulate, but not as much as I should be."* This began a conversation in which a shared meaning of inquiry was established. Through a 1.5 hour conversation driven by the teacher participants with minimal facilitation by the researchers, the teachers arrived at their own definition of inquiry: *"Socially constructing evidence-based meaning of phenomena through intentionally sequenced events."*

## CONCLUSIONS AND IMPLICATIONS

The findings of this study indicate that these practicing teachers' conceptions of inquiry, which has been a central focus of science education reform for decades, were initially unclear. Their conceptions evolved and became more expert-like as they participated in this teacher-driven professional development program. Each has extensive teacher preparation, and, though they each felt inquiry was an important aspect of physics instruction, none of them appeared to hold sophisticated conceptions of inquiry at the beginning of the study.

As these teachers engaged in collaborative discourse, their understanding appeared to evolve, and these teachers came to the realization that their own conceptions of inquiry had been unclear and in need of refinement. These findings raise important questions concerning the preparation of teachers for the physics classroom. Why, in spite of the national efforts to enhance inquiry learning in science, have these teachers found themselves struggling to understand what inquiry teaching and learning is?

These findings suggest that traditional teacher preparation is likely too abstracted from real teaching practice to provide teachers with the learning experiences necessary to develop rich and robust understanding of complex phenomenon such as inquiry-oriented science instruction. It may be that pre-service teachers could benefit greatly from teacher learning that is grounded more firmly in authentic classroom experience and collaborative processing of those experiences. Further, we must consider the advantages of the physics content specific learning experiences for teachers, particularly in light of the fact that traditional teacher preparation too often fails to provide meaningful and sustained learning experiences in pre-service teachers' content areas. And, though it has been shown that supplemental teacher education programs, such as the Learning Assistant model [5], have a positive impact on teachers' preparation, the fact that two of the four teachers in this study were former LAs indicates that more reforms are likely necessary.

In addition to raising questions about what our teachers learn, these finding point to important questions about *how* our teachers learn. What are the implications of what these teachers accomplished together, particularly when considered with respect to what their preparation experiences did not? Perhaps when we consider teacher education and professional development, we might reconsider what our most valuable and effective resources are. Traditionally, we have sought to bring "experts" to our teachers in hopes that they might impart upon them the wisdom that manifests in effective practice. This study suggests that our most valuable resources for teacher growth may be the teachers themselves, drawing on their everyday classroom experiences and working collaboratively toward a greater, shared understanding of the topics that they recognize as central to improving their own practices. We might reconsider our efforts to "give professional development to" our teachers and start thinking about creating learning communities that rely on the professionalism and experience-based expertise of teachers to affect change.

## ACKNOWLEDGEMENTS

We are grateful to the Streamline teachers and NSF DUE #934921.

## REFERENCES

1. R.D. Anderson, Journal of Science Teacher Education **13**(1), 1-12 (2002).
2. M.A. Windschitl, Journal of Research in Science Teaching, **41**(5). 481-512 (2004).
3. N.H. Kang, M.K. Orgill, and K.J. Crippen, Journal of Science Teacher Education **19**, 337-354 (2008).
4. National Research Council, *Inquiry and the NSES* (National Academies Press, Washington DC, 2000).
5. V.K. Otero, S.J. Pollock, and N.D. Finkelstein, Am. J. Phys. **78**, 1218 (2010).